\documentstyle[aps,prl,epsf,rotate]{revtex}
\begin{document}

\twocolumn[\hsize\textwidth\columnwidth\hsize\csname     
@twocolumnfalse\endcsname

\title{Highly Anisotropic Transport in the Integer Quantum Hall Effect
}

\author{W. Pan$^{1,2}$,
H.L. Stormer$^{3,4}$,
D.C. Tsui$^{1}$,
L.N. Pfeiffer$^{4}$, K.W. Baldwin$^{4}$, and
K.W. West$^{4}$}
\address{$^{1}$Department of Electrical Engineering, Princeton
University,
Princeton, New Jersey 08544}
\address{$^{2}$National High Magnetic Field Laboratory, 
Tallahassee, Florida 32310}
\address{$^{3}$Department of Physics
and Department of Applied Physics, Columbia University, New
York,
New York 10027}
\address{$^{4}$Bell Labs, Lucent Technologies, Murray Hill, New
Jersey 07974}

\date{\today}
\maketitle

\begin{abstract}

At very large tilt of the magnetic ($B$) field with respect to the plane of
a two-dimensional electron system the transport in the integer quantum
Hall regime at $\nu$ = 4, 6, and 8 becomes strongly anisotropic. At these
filling factors the usual {\em deep minima} in the magneto-resistance occur
for the current flowing {\em perpendicular} to the in-plane $B$ field
direction but develop into {\em strong maxima} for the current flowing
{\em parallel} to the in-plane $B$ field. The origin of this anisotropy
is unknown but resembles the recently observed anisotropy at
half-filled Landau levels. 

\end{abstract}

\vskip2pc]

Strongly correlated electronic systems often exhibit stripe phases
\cite{emery:prl00}. In two-dimensional electron systems (2DES) 
such a stripe phase is
believed to be at the origin of the recently observed electronic
transport anisotropy at half-fillings of high Landau levels 
\cite{horst:aps93,lilly:prl99a,du:ssc99,pan:prl99,lilly:prl99b,shayegan:physicae99,pan:prl00}. At
Landau level filling factors $\nu$ = 9/2, 11/2, 13/2, etc. the
magneto-resistance is a maximum along one current direction, whereas
it is a minimum when the current direction is rotated by 90$^{\circ}$ within
the plane of the sample. In a purely perpendicular magnetic field ($B$)
the direction of anisotropy is pinned to the crystal lattice \cite{lilly:prl99a,du:ssc99},
but re-orients itself when an in-plane $B$ field ($B_{ip}$) is added
by tilting the sample. At large $B_{ip}$ the easy-axis of anisotropy in the
plane of the sample (the direction of minimum resistance) is {\em always}
perpendicular to $B_{ip}$ \cite{pan:prl99,lilly:prl99b}. 
Although the nature of this anisotropy
remains uncertain, experimental data \cite{horst:aps93,lilly:prl99a,du:ssc99,pan:prl99,lilly:prl99b,shayegan:physicae99,pan:prl00}
and theoretical models
\cite{koulakov:prl96,moessner:prb96,fradkin:prb99,fertig:prl99,rezayi:prl99,simon:prl99,jungwirth:prb99,philip:prl00,fradkin:prl00,maeda:prb00,macdonald:prb00,vonoppen:prl00,cote:prb00} point
to the formation of a unidirectional charge density wave,
often referred to as the ``stripe phase'', or to a state akin to a
liquid crystal phase \cite{fradkin:prb99}. 
A very similar anisotropy is also observed
at $\nu$ = 5/2 and $\nu$ = 7/2 in the second Landau level under large $B_{ip}$
\cite{pan:prl99,lilly:prl99b}.
Modeling \cite{rezayi:prl00} suggests that an electronic anisotropic phase, not
unlike the one at half-fillings of higher Landau levels, has been
induced by the in-plane $B$ field.
 
So far, anisotropy has only been observed at {\em half-filled}
Landau levels. In this letter, we present data that show strong
electronic transport anisotropies at {\em fully filled} Landau levels. They
are created by the very strong in-plane $B$ fields at very large
tilt in the regime of the integral quantum Hall effect (IQHE) at $\nu$ =
4, 6, and 8. The origin of these anisotropies is unknown, although,
phenomenologically, they resemble the anisotropies at half-filled
Landau levels: the magneto-resistance is a minimum when the current is
perpendicular to $B_{ip}$ and a maximum when the current is along $B_{ip}$. A
striped spin density wave phase may be at the origin of these new
observations.

Our sample consists of a 350\AA~wide GaAs quantum well embedded into
Al$_{.24}$Ga$_{.76}$As and delta-doped from both sides at a distance of 490\AA.
The specimen has a size of 5mm $\times$ 5mm and is contacted via eight indium
contacts placed symmetrically around the perimeter. The electron
density is established after illuminating the sample with a red
light-emitting diode at $\sim$ 4.2K and, within limits, the density can be
tuned by exposure time. 
At an electron density of $n = 4.2\times10^{11}$ cm$^{-2}$
two electrical subbands are populated having densities 
$n_0 \sim 3.1\times10^{11}$
cm$^{-2}$ and $n_1 \sim 1.1\times10^{11}$ cm$^{-2}$ as 
determined by Fourier analysis of the
low-field Shubnikov-de Haas oscillations. All angular dependent
measurements are carried out in a dilution refrigerator equipped with
an {\em in-situ} rotator placed inside a 33 Tesla resistive magnet. We
define the axis of rotation as the {\it y}-axis. Consequently, the in-plane
field, $B_{ip}$, extends along the {\it x}-axis when the sample is rotated.
	
We have measured $R_{xx}$ and $R_{yy}$, which differ only in the
in-plane current direction, at more than 10 tilt angles ($\theta$) between
0$^{\circ}$ 
and 90$^{\circ}$. $R_{xx}$ represents the direction for which, under tilt, the
current runs along $B_{ip}$.
Figure 1 shows data at five selected angles, from $\theta$ = 81.1$^{\circ}$ 
to 84.4$^{\circ}$. 
At $\theta$ = 0$^{\circ}$ (not shown) both $R_{xx}$ and $R_{yy}$ vanish
at $\nu$ = 6 as expected for an isotropic quantum Hall state. As $\theta$ is
increased towards 81.1$^{\circ}$, both $R_{xx}$ and $R_{yy}$ remain vanishingly small at
$\nu$ = 6, although the widths of the resistance minima and of the Hall
plateau shrink with increasing $\theta$. Very generally, such an angular
dependence is readily understood for the spin-unpolarized $\nu$ = 6 state.
While the $\nu$ = 6 state always occurs at the same perpendicular magnetic
field, $B_{perp}$, the total magnetic field at tilt angle, 
$\theta$, increases as
$B_{tot}=B_{perp}$/cos($\theta$). Since the electron 
spin experiences $B_{tot}$, the
Zeeman splitting of all Landau levels increases with increasing $\theta$.
This leads to a reduction of the energy gap at $\nu$ = 6 and a shrinking
width and depth (not visible on the linear scale of Fig.~1) of the $R_{xx}$
and $R_{yy}$ minima. Eventually, this leads to a collapse and disappearance
of the $\nu$ = 6 IQHE state. Indeed, at $\theta$ = 83.3$^{\circ}$, 
$R_{xx}$ has turned from a
deep minimum into a {\em strong peak} and the usual Hall plateau has
vanished. Therefore, the disappearance of $R_{xx}$ can be rationalized as
the closing of the $\nu$ = 6 energy gap. However, very surprisingly, the
electrical transport turns out to be {\em strongly anisotropic}. In contrast
to $R_{xx}$, which shows a strong {\em maximum} at 
this angle and filling factor,
$R_{yy}$, continues to shows a strong {\em minimum} at $\nu$ = 6. 
Just as in the case
of half-fillings \cite{pan:prl99,lilly:prl99b} 
the easy-axis of this anisotropy at
full-filling factors is perpendicular to $B_{ip}$. The direction of
anisotropy is not dependent on the orientation of the crystallographic
axis with respect to the in-plane field, as we determined by
performing the same experiments on the same specimen mounted in a
configuration rotated by 90$^{\circ}$ about the sample normal. Furthermore,
none of the resistance measurements showed any hysteresis as a
function of the sweep direction of the $B$ field. Finally, in the
anisotropic regime, the generally strong Hall plateau at $\nu$ = 6
disappears for both directions of current.

This is the first time that such an anisotropy has been
observed in a state as robust as an IQHE state. To learn more about
this anisotropic state we perform $T$-dependent studies of $R_{xx}$
and $R_{yy}$. For comparisons, we choose $\theta$ = 81.1$^{\circ}$, 
where the electronic
transport is isotropic, and $\theta$ = 83.3$^{\circ}$, 
where transport is strongly
anisotropic. in Figure 2a and Figure 2b 
we show three representative traces of $R_{xx}$
and $R_{yy}$. At $\theta$ = 81.1$^{\circ}$, $R_{xx}$ and $R_{yy}$ 
exhibit the usual activated
behavior: the value of both resistances increases with increasing
$T$. On the other hand, at $\theta$ = 83.3$^{\circ}$, 
$R_{xx}$ and $R_{yy}$ behave
oppositely: $R_{xx}$ decreases whereas $R_{yy}$ increases with increasing
$T$. The $T$-dependencies are quantified in Figure 2c
and Figure 2d, where $R_{xx}$ and $R_{yy}$ are shown on Arrhenius plots. 
At $\theta$ =
81.1$^{\circ}$, $R_{xx}$ and $R_{yy}$ show well-behaved 
activated behavior yielding a
single energy gap of $\Delta \sim$ 1K for 
both current directions \cite{footnote1}. On
the other hand, the data for $R_{xx}$ and $R_{yy}$ 
at $\theta$ = 83.3$^{\circ}$ show no longer
activated behavior. $R_{xx}$ and $R_{yy}$ appear to start from similar values at
high temperature but then diverge from each other roughly
exponentially with exponents of similar magnitude but {\em opposite sign}.
At the lowest temperatures both resistances assume an approximately
$T$-independent behavior. This dependence is qualitatively the same as
the $T$-dependence of the anisotropic state 
at $\nu$ = 9/2 \cite{du:ssc99,fradkin:prl00}. 

The remarkable anisotropy found in the IQHE is not limited to
the $\nu$ = 6 state. Similar anisotropies are observed at filling factors
$\nu$ = 4 and $\nu$ = 8. Figure 3 shows the $\nu$ = 8 
and $\nu$ = 4 anisotropy in the
same sample at slightly different densities, tuned by applying
different doses of light. We have not performed a systematic
study of these states.
 
The cause of the anisotropy at integral quantum Hall states is
unknown. Before speculating about the origin of this new phenomenon it
is instructive to consider in more detail the single particle states
in this two-electric subband specimen. Figure 4a shows the usual Landau
fan diagram for a density of 4.2 $\times 10^{11}$ cm$^{-2}$. 
The Zeeman splitting is
enhanced by a factor of 10 to be visible. The position of the Fermi
level, $E_f$, is indicated by a heavy line. Clearly, in the vicinity of
$\nu$ = 4, 6 and 8, Landau levels from both electric subbands contribute and
$E_f$ jumps between levels of different origin. Using such a simple
single-particle picture and a 2DES of zero thickness with densities
appropriate for the data of Figs.~1 and 3, one would expect the gaps at
$\nu$ = 4, 6 and 8 to close at $\theta$ = 83.4$^{\circ}$, 
88.4$^{\circ}$ and 88.3$^{\circ}$, respectively.
These values differ from experiment, especially in the case of the $\nu$ =
6 and $\nu$ = 8 states.
 
The discrepancy is largely the result of the neglect of exchange and
of the thickness of the wave function.
In the remainder we focus on the state at $\nu$ = 6, which we studied
most extensively and which shows the strongest anisotropy in
experiment. We expect similar arguments to hold for $\nu$ = 4 and $\nu$ = 8.
Figure 4b shows the result of a self-consistent
local-density-approximation calculation \cite{jungwirth} 
performed for a density $n = 4.2\times 10^{11}$ cm$^{-2}$
at a filling factor $\nu$ = 6 as a function of $B_{ip}$. The
gap at $\nu$ = 6 (shaded region) undergoes strong variations, comes almost
to a close at $B_{ip} \sim 2.5$T (not shown), 
and vanishes at $B_{ip} \sim 18.5$T due to
level crossing. The experimental value of $B_{ip}$ for the strong
anisotropy is $\sim$25T. However, we consider the theoretical result of
$\sim$18.5T to be sufficiently close to $\sim$25T to attribute the disappearance
of the energy gap at $\nu$ = 6 in Fig.~1 to the crossing of spin-split
Landau levels originating from different electrical subbands
({\it i} = 1, 2).
This provides a rational for the appearance of novel features in the
data at this filling factor and angle. However, none of such level
crossing considerations can explain the observed {\em anisotropy}, which
represents the remarkable finding in our data. The origin of this
phenomenon must be the result of correlated electron behavior.

Previously, large electrical anisotropies have only been observed at
half-filled Landau levels \cite{horst:aps93,lilly:prl99a,du:ssc99,pan:prl99,lilly:prl99b,shayegan:physicae99,pan:prl00}.
It is believed that there the
electron system spontaneously breaks into striped domains of
alternating filling factors such as $\nu$ = 4 and $\nu$ = 5 around $\nu$ = 9/2
\cite{koulakov:prl96,moessner:prb96,fradkin:prb99,fertig:prl99,rezayi:prl99,simon:prl99,jungwirth:prb99,philip:prl00,fradkin:prl00,maeda:prb00,macdonald:prb00,vonoppen:prl00,cote:prb00}.
Given the similarity of the observed properties of the
anisotropic phases around $\nu$ = 9/2 and $\nu$ = 6 one might speculate on a
similar underlying striped geometry. The driving force behind the
phase separation in the $\nu$ = 9/2 case is exchange. The energetic gain
from breaking into domains of $\nu$ = 4 and $\nu$ = 5 is counteracted by a
strong electrostatic cost for creating an inhomogeneous charge
distribution. This is the reason for the formation of very narrow
stripes of $\nu$ = 4 and $\nu$ = 5 states, which are only a few magnetic
lengths wide. A phase, consisting of stripes around $\nu$ = 6, would carry
a much smaller, electrostatic burden.

At the point of collapse of the $\nu$ = 6 energy gap in Fig.~4b two
electronic configurations are degenerate. At $B_{ip}$ smaller than the
level crossing in Fig.~4b the electrons occupy three spin-unpolarized
levels emanating from the lowest three Landau levels ({\it N} = 0, 1,
and 2) of the
lower electronic subband, {\it i} = 1 \cite{footnote2}. 
(Note, an earlier anti-crossing at
$B_{ip} \sim 2.5$T exchanges states {\it i} = 1, {\it N} = 2 and {\it
i} = 2, {\it N} = 0). The total system is
spin-unpolarized (3 spin-up, 3 spin-down). At $B_{ip}$ larger than the
level crossing in Fig.~4b the electrons occupy only two
spin-unpolarized levels emanating from the lowest two Landau levels
({\it N} = 0, 1) of the lower electronic subband ({\it i} = 1). 
In addition, they
occupy the spin-up states (solid lines) of two levels emanating from
the {\it i} = 1, {\it N} = 2 and the {\it i} = 2, {\it N} = 0 states. 
There, the total system is
{\em partially spin-polarized} (4 spin-up, 2 spin-down). In the vicinity of
the level crossing in Fig.~4b, a phase separation of the electronic
system \cite{giuliani:prb85,yalagadda:prb91} 
into spin-unpolarized and partially spin-polarized
domains may occur driven by exchange. A very small gain in exchange
energy may suffice, since the charge density in both configurations is
identical and, to first order, there is no associated electrostatic
cost. 

Such a pattern resembles the pattern of a spin-density wave, SDW. The
existence of an in-plane magnetic field and the so-induced coupling of spin and
orbital motion will energetically favor a given orientation of the
stripes with respect to $B_{ip}$. The resulting stripe phase of alternating
IQHE configurations is bound to have one-dimensional edge-states along
its interface between neighboring domains, which carry the electric
current in a highly anisotropic fashion. This transport pattern would
be analogous to the pattern invoked in the stripe phases that are
believed to form at half-fillings of Landau levels, such as $\nu$ = 9/2
and 13/2 and believed to be responsible for the anisotropic electronic
behavior. However, without the application of other experimental
techniques and without a detailed theoretical investigation this
picture remains speculative.

In summary, we have observed strongly anisotropic transport
under high in-plane magnetic field in the regime of the IQHE in a
quantum well sample with two occupied electrical subbands.
Phenomenologically, the data have much in common with the previously
discovered anisotropy at half-fillings of high Landau levels.
From a simple level crossing picture we conjecture that
a novel striped spin-density wave may be at the origin of this
phenomenon.

We would like to thank E. Palm and T. Murphy for experimental
assistance, E. P. De Poortere, S. P. Shukla, and E. Tutuc for the help
in numerical calculation, and N. Bonesteel, R. R. Du, A.H. MacDonald,
N. Read, and K. Yang for useful discussion. We are indebted to T.
Jungwirth for providing the energy level scheme of Figure 4b. A
portion of this work was performed at the National High Magnetic Field
Laboratory, which is supported by NSF Cooperative Agreement No.
DMR-9527035 and by the State of Florida. D.C.T. and W.P. are supported
by the DOE and the NSF.

\vspace*{-0cm}

\begin{figure}[t]
\epsfxsize=2.0in
\centerline{
\epsffile{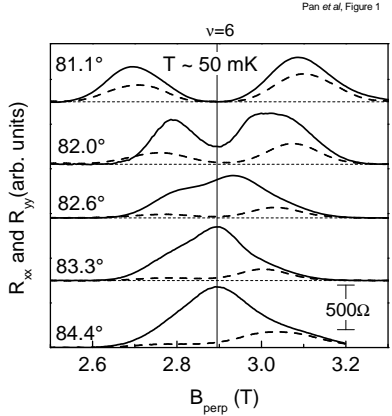}}

\vspace*{-0cm}

\caption{
$R_{xx}$ (solid lines) and $R_{yy}$ (dashed lines) around
$\nu$ = 6 at $T \sim$ 50~mK, $I$ = 10~nA and for five tilt
angles, from $\theta$ = 81.1$^{\circ}$ to $\theta$ = 84.4$^{\circ}$. 
For $R_{xx}$
the current runs along
the in-plane magnetic field.
$n = 4.2 \times 10^{11}$ cm$^{-2}$.
}
\end{figure}

\begin{figure}[t]
\epsfxsize=2.5in
\centerline{
\epsffile{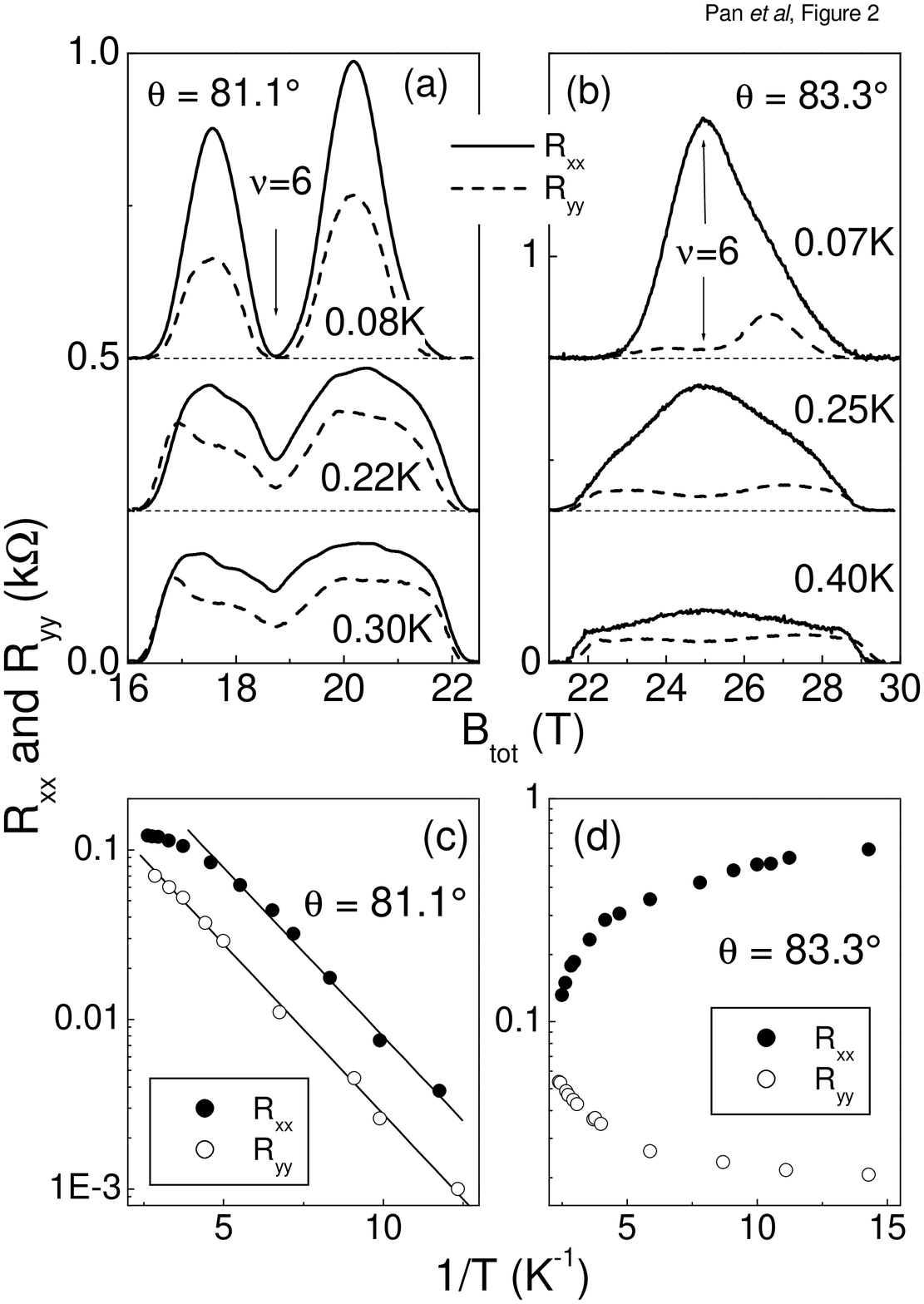}}

\vspace*{-0cm}

\caption{
Panels (a) and (b): Temperature dependence of $R_{xx}$ (solid lines) and
$R_{yy}$ (dashed lines) at tilt angles $\theta$ = 81.1$^{\circ}$ (a) 
and 83.3$^{\circ}$ (b) and
three different temperatures each. Traces are shifted vertically for
clarity. The position of the $\nu$ = 6 filling factor is indicated. Panels
(c) and (d): Corresponding Arrhenius plots for $R_{xx}$ and $R_{yy}$ 
at $\nu$ = 6 at
the tilt angles of the panels above. The straight lines in panel (c)
are a linear fit to the data \protect\cite{footnote1}. 
The energy gap is $\sim$ 1K for $R_{xx}$ and $R_{yy}$.
}
\end{figure}

\begin{figure}[t]
\epsfxsize=2.0in
\centerline{
\epsffile{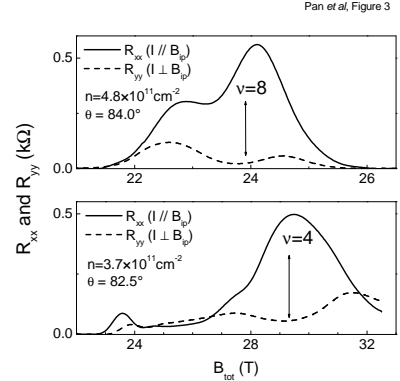}}

\vspace*{-0cm}

\caption{
(a) Anisotropic transport around $\nu$ = 8 at $\theta$ =
84.0$^{\circ}$. 
(b) Anisotropic
transport around $\nu$ = 4 at $\theta$ = 82.5$^{\circ}$. 
The sample densities are slightly
different from Fig.~1 and have been adjusted by different light
exposure. For $R_{xx}$ the current runs along the in-plane magnetic field.
}
\end{figure}

\begin{figure}[t]
\epsfxsize=2.0in
\centerline{
\epsffile{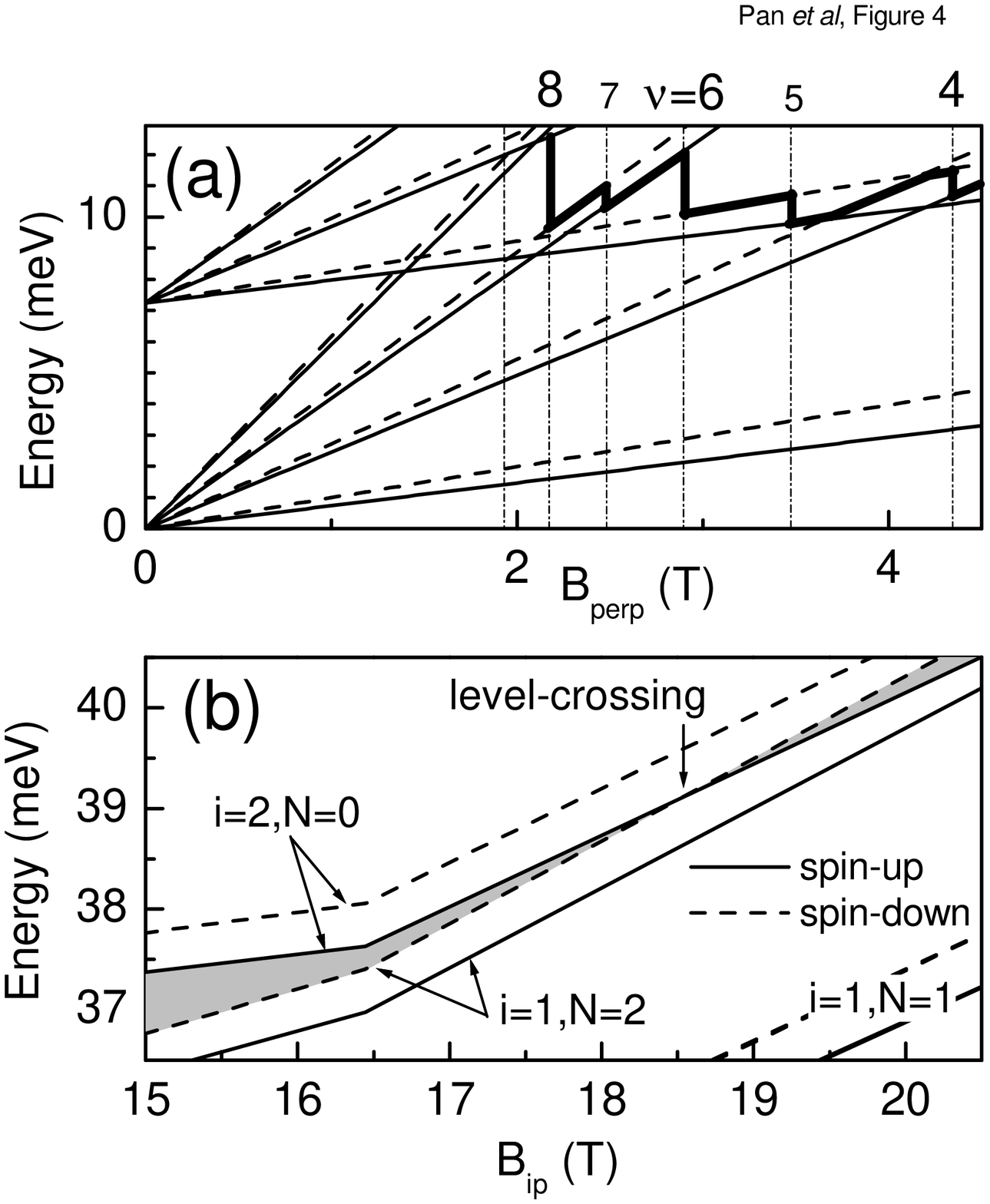}}

\vspace*{-0cm}

\caption{
Panel (a): Simple Landau fan diagram for the two-electric subband sample of
density $n = 4.2\times 10^{11}$ cm$^{-2}$. The Zeeman splitting is enhanced by a
factor of 10 to be visible. The position of the Fermi level is
indicated by heavy line. Panel (b): Result of self-consistent
local-density-approximation calculation \protect\cite{jungwirth} 
at $\nu$ = 6 and as a
function of in-plane magnetic field, $B_{ip}$, at the same density as panel
(a). The electric subband index, {\it i}, and the Landau level index,
{\it N}, are
indicated. The shaded area represents the gap at $\nu$ = 6.
}
\end{figure}


\begin{references}
\bibitem{footnote1}
$R_{xx}$ is about three times larger than $R_{yy}$ at the same temperature. This
level of initial anisotropy is not uncommon in 2DES, whether they are
patterned or cleaved into squares. It is probably due to small density
variations leading to current inhomogeneities. However, it cannot be
the origin of the strong anisotropy since the strong anisotropy
observed at $\nu$ = 4, 6, and 8 is pegged to 
$B_{ip}$ and not to any particular sample
direction.
\bibitem{footnote2}
Strictly speaking, the separation into subband levels and Landau
levels only holds at zero tilt angle. At non-zero tilt they are mixed
and one can use this nomenclature only approximately.
\end{references}
\end{document}